# On a charge conserving alternative to Maxwell's displacement current

Alan M Wolsky[1,2 a)b)]
[1] Argonne National Laboratory, 9700 South Cass Ave., Argonne IL 60439
[2] 5461 Hillcrest Ave., Downers Grove, IL 60515

[a)] E- mail: AWolsky@ANL.gov
[b)] E- mail: AWolsky@ATT.net

**Abstract**. Though sufficient for local conservation of charge, we show that Maxwell's displacement current is not necessary. An alternative to the Ampere-Maxwell equation is exhibited and the alternative's electric and magnetic fields and scalar and vector potentials are expressed in terms of the charge and current densities. The alternative describes a theory in which action is instantaneous and so may provide a good approximation to Maxwell's equations where and when the finite speed of light can be neglected. The result is reminiscent of the Darwin approximation which arose from the study classical charged point particles to order $(v/c)^2$ in the Lagrangian. Unlike Darwin, this approach does not depend on the constitution of the electric current. Instead, this approach grows from a straightforward revision of the Ampere Equation that enforces the local conservation of charge.

# I. INTRODUCTION

When, after more than a few decades, the author returned to Maxwell's equations, his attention was held by a formerly unnoticed imprecision in the textbook treatment of Maxwell's displacement current.

All celebrate its addition to the right hand side of Ampere's Law as Maxwell's masterstroke and most texts motivate its addition by discussing the inadequacy of Ampere's Law where and when the current density's divergence is not zero.

Indeed, the student cannot be other than delighted by how the addition of the displacement current generalizes Ampere's Law and how that generalization entails the local conservation of charge. It hardly occurs to him or her to ask if the addition was necessary. After all, some books say it is and others do not comment. Instead, they assert that the addition completes Maxwell's Equations and it remains only to find their solutions for a variety of materials and circumstances. This hardly invites attention to possible alternatives.

The imprecision is simply stated; textbooks do not distinguish between the necessity and sufficiency of Maxwell's displacement current to ensure local charge conservation. We will show that, though it is sufficient, Maxwell's displacement current is not necessary. There are other ways to ensure local charge conservation.

There are two reasons to dwell on this point. The first is to emphasize to graduate students that, sometimes, there is more than one way to skin a cat – in this case to reconcile Ampere's Law with local charge conservation – a lesson that is most convincingly taught by example. The second is that by exploring alternate theories, one might find that one - even if less appealing (e.g., less comprehensive or less accurate) than another - has equations that are more amenable to solution. And so, these might usefully approximate the more appealing theory. Indeed, we shall find that this is so for the example developed on the following pages.

The rest of this paper is organized as follows. Section II presents the promised alternative to the Ampere-Maxwell Equation. Logically, this is the punch line. However, what it entails is not immediately obvious. And so, the alternative's similarities to and differences from Maxwell's Equations are the subject of the rest of this paper.

Section III shows how the magnetic field depends on the current when the alternative to the Ampere-Maxwell Equation is adopted. This dependence draws attention to the vector potential and so the potentials, vector and scalar, are discussed in Section IV where the alternative to Maxwell is shown to be gauge invariant. Section V concerns itself with the vector potential and the electric field. This appears to be more difficult to calculate than in the static case. Happily this difficulty is shown to be only apparent and not real.

One finds that indeed, in some circumstances, the alternative may provide a useful approximation to Maxwell's Equations. To put all this in context, a brief section recalling the history of related matters follows. Then, we summarize and conclude.

## II.  AN ALTERNATIVE TO MAXWELL'S EQUATIONS

For the sake of ready reference, we begin by recalling the four "pre-Maxwell Equations"

$$div[\mathbf{E}] = \rho/\varepsilon_0 \qquad (\text{II}.1)$$

$$div[\mathbf{B}] = 0 \qquad (\text{II}.2)$$

$$\mathbf{curl}[\mathbf{E}] = -\partial_t \mathbf{B} \qquad (\text{II}.3)$$

$$\mathbf{curl}[\mathbf{B}] = \mu_0 \mathbf{J} \qquad (\text{II}.4)$$

where $\rho$ denotes the total (free and polarization) charge density and $\mathbf{J}$ denotes the total (free and magnetization) current density.

And we must not forget, Eq.II.5, the fifth pre-Maxwell Equation; it expresses the local conservation of charge, a requirement that will figure prominently in our considerations.

$$div[\mathbf{J}] = -\partial_t \rho \qquad (\text{II}.5)$$

Indeed, Eq.II.5 does more than assert current conservation. It relates the current density that is the source of the static magnetic field to the charge density that is a source of the static electric field.

The first four pre-Maxwell Equations describe situations in which the divergence of the current density, $\mathbf{J}$, is zero. Of course, these situations are not the only ones of interest. One also wants to know what happens where and when the charge density is changing. The pre-Maxwell Equations do not provide an answer. The divergence of the left hand side of Ampere's Equation, Eq.II.4. is always zero but the divergence of its right hand side is not. Thus the question, "How should Ampere's Equation be revised to describe all situations?" As is well known, Maxwell's answer was to add the term $\partial_t(\varepsilon_0 \mathbf{E})$ and thereby arrive at what is often now often called the Ampere-Maxwell Equation

$$\mathbf{curl}[\mathbf{B}] = \mu_0 \left( \mathbf{J} + \partial_t(\varepsilon_0 \mathbf{E}) \right) \qquad (\text{II}.6)$$

This does the job because we believe in the local conservation of charge, as expressed by Eq.II.5. And so the vanishing divergence of the left-hand side of Eq.II.6 is matched to the vanishing divergence of the right-hand side. Alternatively, one could say that Eq.II.6 entails the local conservation of charge, something we are predisposed to believe. In summary, the Ampere-Maxwell Equation is sufficient to guarantee charge conservation.

But, is it necessary? Is there a different revision to the Ampere Equation that would achieve the same result? Well yes, there is. That different revision comes to mind when one recalls that the divergence of the gradient is the Laplacian and so one has Eq.II.7.

$$div_x \left[ \nabla_x \frac{1}{4\pi \|\mathbf{x} - \mathbf{y}\|} \right] = \Delta_x \frac{1}{4\pi \|\mathbf{x} - \mathbf{y}\|} = -\delta^3[\mathbf{x} - \mathbf{y}] \qquad (\text{II}.7)$$

Thus Eq.II.8 presents a different way to generalize Ampere's Equation.

$$\mathbf{curl}\big[\mathbf{B}[t,\mathbf{x}]\big] = \mu_0 \left( \mathbf{J}[t,\mathbf{x}] - \nabla_x \iiint_{all\,space} d^3y \, \frac{\partial_t \rho[t,\mathbf{y}]}{4\pi \|\mathbf{x}-\mathbf{y}\|} \right) \qquad (\text{II.8})$$

After taking the divergence of both sides, one confirms that Eq.II.8 entails the local conservation of charge and thus our addition presents an alternative to Maxwell's displacement current. Maxwell called the sum of the conduction current and his displacement current the true current and denoted it by **C**.[1] We will denote the analogous quantity in Eq.II.8 by $\mathbf{J}^{ATM}$ where ATM stands for Alternative to Maxwell.

As an aside, we note that we could have adopted a superficially different alternative. We could have asserted that only the transverse or solenoidal part of the current density belongs on the right hand side of Ampere's equation, as shown in Eq.II.9

$$\mathbf{curl}\big[\mathbf{B}[t,\mathbf{x}]\big] = \mu_0 \left( \mathbf{J}[t,\mathbf{x}] + \nabla_x \iiint_{all\,space} d^3y \, \frac{div\big[\mathbf{J}[t,\mathbf{y}]\big]}{4\pi \|\mathbf{x}-\mathbf{y}\|} \right) \qquad (\text{II.9})$$

As desired, the divergence of each side of this equation vanishes and as desired it reduces to Ampere's Law when and where the divergence of the current density vanishes. However, this equation by itself does not effect the unification of electricity and magnetism. Nothing in it relates the current, that is the source of **B** to charge that is the source of **E**. To effect the needed unification, one must separately assert local conservation of charge, Eq.II.5. This separate assertion seems unobjectionable and so we see no significant difference between these approaches, which is why we called their difference superficial.

Thus, we have arrived at a charge conserving alternative to the Ampere-Maxwell Law. Maxwell's displacement current is sufficient for local charge conservation but it is not necessary.[2]

While most texts are mum on this point, several otherwise excellent ones simply err. Here are five examples, each drawn from a different, widely used text. "We needed the new term [i.e., *Maxwell's displacement current*] to make the relation between current and magnetic field consistent with the continuity equation, in the case of conduction currents changing in time."(Purcell).[3] "Why the displacement current is necessary." (Pollack and Stump)[4] "The need for the addition of this term [*Maxwell's displacement current*] to produce a solenoidal net current vector was recognized by Maxwell."(Panofsky and Phillips)[5] "The additional term [i.e., *Maxwell's displacement current*] is necessary to fulfill the continuity equation."(Greiner)[6] "Maxwell deduced that a changing electric field must produce a magnetic field."(Garg)[7] Happily, one text (Griffiths) is different; it alerts its readers that "… theoretical convenience and aesthetic consistency are only *suggestive*-there might, after all, be other ways to doctor up Ampere's Law..."[8] The doctoring up, sans pejorative, is provided by Eq.II.8.

Of course, the contemporary mind will look for something wrong with Eq.II.8. It is apparent that it involves an integral over an extended region in space at a single time. And so the question will be "in which inertial frame does this equation describe the physics?" The reply is straightforward. The equation describes charges, currents and fields in material (e.g., aluminum, copper, iron, polyethylene, air) that is everywhere at rest in the same inertial frame. As will be mentioned later, one could claim a wider domain of applicability but for our (pedagogical) purpose there is no point to further elaboration.

The next question may well be, "Is this a field theory, like Maxwell's, or Action-At-A-Distance or some kind of hybrid?" And esthetics aside, the root question is, "How do the solutions of the ATM equations differ from the solutions of Maxwell's Equations?"

The following section is devoted to the first part of the answer. The magnetic field's dependence on the current will be presented and discussed.

Before doing so, one mathematical aside is in order. When comparing equations and their solutions we must do so for the same distributions of charge and current and the same boundary conditions. While discussing the solutions of Maxwell's equations, it is always assumed that charges and currents vanish at infinity so that surface integrals at infinity may be neglected. It is also assumed that fields vanish at infinity. This is suggested by measurements that show mutual influence diminishes as separation increases, as specified by Coulomb's and Biot-Savart's Laws which are particular solutions of Poisson's Equation.[9] We will make the same assumptions when discussing the solutions of the ATM Equations.

**III. How the ATM magnetic field depends on the current**

Here we ask for the solution of Eq.II.3 and Eq.II.9. The former denies the existence of magnetic poles and the latter is ATM's alternative to the Ampere-Maxwell Equation. The answer is a bit of a surprise. It is simply the Biot-Savart Law which appears below on the right hand side of Eq.III.1.

$$\mathbf{B}^{ATM}[t,\mathbf{r}] = \mu_0 \iiint_{all\,space} \frac{d^3 r_s}{4\pi} \frac{\mathbf{J}[t,\mathbf{r}_s] \times (\mathbf{r} - \mathbf{r}_s)}{\|\mathbf{r} - \mathbf{r}_s\|^3} \qquad (III.1)$$

The surprise comes from the circumstance that physicists first meet this expression in a textbook chapter devoted to magnetostatics, after which the Biot-Savart Law is forgotten.[10] Nonetheless, as shown in Eq.III.2, straightforward calculation confirms that Biot-Savart solves Eq.II.9, the Ampere-ATM Equation.

$$\mathbf{curl}_r \left[ \mu_0 \iiint_{all\,space} \frac{d^3 r_s}{4\pi} \frac{\mathbf{J}[t,\mathbf{r}_s] \times (\mathbf{r} - \mathbf{r}_s)}{\|\mathbf{r} - \mathbf{r}_s\|^3} \right] = \mu_0 \left( \mathbf{J}[t,\mathbf{r}] + \nabla_r \iiint_{all\,space} \frac{d^3 r_s}{4\pi} \frac{div[\mathbf{J}[t,\mathbf{r}_s]]}{\|\mathbf{r} - \mathbf{r}_s\|} \right) \qquad (III.2)$$

This result is often unrecorded though it is sometimes stated without comment, other than to dismiss the "$div[\mathbf{J}]$" term because the chapter's concern is magnetostatics.[11,12] We do not dismiss; we take this term seriously. It shows that the Biot-Savart expression is not limited to statics and, in particular, it is a solution of the Ampere-ATM Equation, a point nicely illustrated by a recent article in this journal. [13] For discussion of generalizations of the Biot-Savart Law in light of Jefimenko's [14] expression for the fields that solve Maxwell's equations, see Griffiths and Heald, who however do not take account of Eq.III.2 [15]

Also at hand is a simple proof that this expression requires the absence of magnetic poles. One just notes that the right hand side of Eq.III.1 can be expressed as the curl of a vector, as shown in Eq.III.3.

$$\mathbf{B}^{ATM}[t,\mathbf{r}] = \mu_0 \iiint_{all\ space} \frac{d^3 r_s}{4\pi} \frac{\mathbf{J}[t,\mathbf{r}_s] \times (\mathbf{r}-\mathbf{r}_s)}{\|\mathbf{r}-\mathbf{r}_s\|^3} = \mathbf{curl}_r \left[ \mu_0 \iiint_{all\ space} \frac{d^3 r_s}{4\pi} \frac{\mathbf{J}[t,\mathbf{r}_s]}{\|\mathbf{r}-\mathbf{r}_s\|} \right] \quad (\text{III.3})$$

The argument of the **curl** is simply a vector potential for $\mathbf{B}^{ATM}$. We will call it the Biot-Savart vector potential and denote it by $\mathbf{A}^{ATM}_{B-S}$.

Finally, we note that the Biot-Savart expression is a particular solution of the ATM equations, not the most general. The general solution is the sum of this particular solution and the solutions of the homogeneous equations, $div[\mathbf{B}] = 0$ and $\mathbf{curl}[\mathbf{B}] = 0$. Though they exist, the homogeneous solutions do not tend to zero as one approaches infinity. One solution is constant and all others grow as one approaches infinity, not the behavior observed in the lab. Thus we rule out these unphysical solutions with the same boundary condition - magnetic fields should diminish as one goes further and further from currents – as is used with Maxwell's Equations.

Consistent with the fact that the Ampere-ATM Equation involves an integral over all space at a same time, $\mathbf{B}^{ATM}$ depends on the currents at the same time and not the retarded (or advanced) time. As such, this is a theory in which the magnetic field propagates instantaneously, not at the speed of light.

Of course, one also wants to know about the electric field. It turns out that a useful answer only comes after a substantially more difficult inquiry. To prepare for it, we will first turn to the relation of the fields to the scalar and vector potentials, in particular to their gauge invariance.

**IV The potentials and their gauge invariance**

As already stated by Eq.II.2, there are no magnetic poles and so we can write $\mathbf{B}^{ATM} = \mathbf{curl}[\mathbf{A}^{ATM}]$. And just as with Maxwell's equations, we deduce from Faraday's Law, $\mathbf{curl}[\mathbf{E}^{ATM} + \partial_t \mathbf{A}^{ATM}] = 0$, that there is a scalar, $\varphi^{ATM}$, such that $\mathbf{E}^{ATM} = -\nabla \phi^{ATM} - \partial_t \mathbf{A}^{ATM}$. When expressed in terms of the potentials, the ATM Equations are

$$-\nabla^2 \phi^{ATM} - \partial_t div[\mathbf{A}^{ATM}] = \rho/\varepsilon_0 \quad (\text{IV.1})$$

$$\nabla div[\mathbf{A}^{ATM}] - \nabla^2 \mathbf{A}^{ATM} = \mu_0 \mathbf{J}^{ATM} \quad (\text{IV.2})$$

Now, two things are clear. First, ATM is gauge invariant that is to say that if a pair of potentials, $\phi^{ATM}$ and $\mathbf{A}^{ATM}$, satisfy these equations then so do many other pairs, each of the form $\phi^{ATM} - \partial_t g$ and $\mathbf{A}^{ATM} + \nabla g$, and all these pairs give the same fields, $\mathbf{E}^{ATM}$ and $\mathbf{B}^{ATM}$. Second, by changing the gauge, one changes only the divergence of the vector potential, not its curl. In this respect, ATM and Maxwell are the same.

To illustrate their variety, two different gauges are defined below.[16] The first is the Biot-Savart Gauge, already mentioned in Section III where we observed that the Biot-Savart expression can be written as the curl of a vector. This vector potential has the same functional dependence on the currents whether or not the currents are divergence free. The second is the widely used Coulomb Gauge. This gauge's scalar potential depends on the charges in the same way whether or not the charge density is time dependent. (When the divergence of the current density vanishes everywhere, these two gauges, Biot-Savart and Coulomb, are the same.). Each gauge has its own heuristic value, which value depends on the task at hand. Below, we present the definition of each followed by the ATM Equations that govern it.

$$\text{Biot-Savart} \qquad div\left[\mathbf{A}_{B-S}[t,\mathbf{r}]\right] \equiv \mu_0 \iiint_{all\ space} \frac{d^3 r_s}{4\pi} \frac{div\left[\mathbf{J}[t,\mathbf{r}_s]\right]}{\|\mathbf{r}-\mathbf{r}_s\|} \qquad (\text{IV.3a})$$

$$-\nabla^2 \varphi_{B-S}^{ATM} = \frac{1}{\varepsilon_0}\left(\rho - \left(\iiint_{all\ space} \frac{d^3 r_s}{4\pi}\left\{\frac{c^{-2}\partial_t^2 \rho[t,\mathbf{r}_s]}{\|\mathbf{r}-\mathbf{r}_s\|}\right\}\right)\right) \qquad (\text{IV.3b})$$

$$-\nabla^2 \mathbf{A}_{B-S}^{ATM} = \mu_0 \mathbf{J} \qquad (\text{IV.3c})$$

$$\text{Coulomb} \qquad div\left[\mathbf{A}_C[t,\mathbf{r}]\right] \equiv 0 \qquad (\text{IV.4a})$$

$$-\nabla^2 \varphi_C^{ATM} = \rho/\varepsilon_0 \qquad (\text{IV.4b})$$

$$-\nabla^2 \mathbf{A}_C^{ATM} = \mu_0\left(\mathbf{J} + \nabla\left(\iiint_{all\ space} \frac{d^3 r_s}{4\pi}\left\{\frac{div\left[\mathbf{J}[t,\mathbf{r}_s]\right]}{\|\mathbf{r}-\mathbf{r}_s\|}\right\}\right)\right) \qquad (\text{IV.4c})$$

Our immediate task is to illuminate the similarity and the difference between ATM and Maxwell. For this, the Coulomb Gauge is particularly helpful. The scalar potential satisfies Poisson's equation in both Maxwell's system and the ATM.

$$-\nabla^2 \phi_C^{ATM} = \rho/\varepsilon_0 = -\nabla^2 \phi_C^{Maxwell} \qquad (\text{IV.5})$$

and so the familiar Coulomb potential appears in both cases. However, the two systems differ in their equation for the vector potential. The second derivative with respect to time appears in Maxwell's Equations and it is absent in the ATM.

$$-\nabla^2 \mathbf{A}_C^{ATM} = \mu_0 \mathbf{J}^{ATM} = c^{-2}\partial_t^2 \mathbf{A}_C^{Maxwell} - \nabla^2 \mathbf{A}_C^{Maxwell} \qquad (\text{IV.6})$$

This difference in the equations entails the difference in their solutions, Eqs.IV.8, which show retarded or advanced time in the solutions of Maxwell's Equations and the simultaneous time in ATM. Thus both ATM's electric and magnetic fields propagate instantaneously.

$$\mathbf{A}_C^{Maxwell}[t,\mathbf{r}] = \iiint_{all\ space} d^3 r_s \frac{\mu_0 \mathbf{J}^{ATM}\left[t \pm c^{-1}\|\mathbf{r}-\mathbf{r}_s\|, \mathbf{r}_s\right]}{4\pi\|\mathbf{r}-\mathbf{y}\|} \qquad (\text{IV.7a})$$

$$\mathbf{A}_C^{ATM}[t,\mathbf{r}] = \iiint_{all\ space} d^3 r_s \frac{\mu_0 \mathbf{J}^{ATM}[t,\mathbf{r}_s]}{4\pi\|\mathbf{r}-\mathbf{r}_s\|} \qquad (\text{IV.7b})$$

That difference reflects the difference between the Ampere-Maxwell Equation, which includes the displacement current, $\varepsilon_0 \partial_t \mathbf{E}$, and the Ampere-ATM Equation which includes the simultaneous rate of change of the charge density, $\partial_t \rho$. One sees that ATM is the limit of Maxwell as $c^{-1}\|\mathbf{r}-\mathbf{r}_s\|$ vanishes. And so we expect ATM to approximate Maxwell when the currents barely change between the present and the retarded or advanced time. For example, when the charge and current densities are periodic with period *T*, we expect the solutions of ATM to approximate the solutions to Maxwell when $\|\mathbf{r}-\mathbf{r}_s\|/cT$ is small. The importance of this regime is apparent to anyone who depends on electric power delivered at 50-60Hz.

In summary, we identified a charge conserving Alternative to Maxwell and, as the result of inquiry, we now find that its acronym, ATM, may also stand for Approximation to Maxwell.

While reading the foregoing, a thought might claim the reader's attention. He or she may recall a subtle feature of Coulomb Gauge in Maxwell's equations that does not appear in the Coulomb Gauge in the ATM. In both, the scalar potential at a point and time depends on the charge density elsewhere and at the same time. In the ATM, this also true of the vector potential. Influence propagates instantaneously. In Maxwell's equations, it does not. By virtue of the appearance of the second time derivative on the right hand side of Eq.IV.6, Maxwell's equations allow propagation of influence at finite speed. In fact, Maxwell's equations require it while still allowing the appearance of instantaneous propagation in the scalar potential of the Coulomb gauge. The detailed description of how Maxwell's equations manage this was long ago presented for sinusoidal sources by Brill and Goodman [17] and then the for arbitrary time dependence by Gardiner and Drummond [18]. Recently Wundt and Jentschura [19] and Frenkel and Rácz [20] claim to have improved earlier expositions. (Jackson [21], Rohrlich [22, 23] and Heras [24] have also commented on this topic and tangentially related issues, as well as their bearing on the reality of Maxwell's displacement current[25].) Those interested in reconciling causality with the Coulomb gauge version of Maxwell's equations may also be interested in the like reconciliation in the superluminal gauges discussed by Yang. [26] Compared with Maxwell's equations, "what you see is what you get" in the ATM, instantaneous Coulomb gauge potentials and instantaneous fields.

**V. On the tractability of ATM as an approximation**

The usual approximations to Maxwell's Equations, magneto-quasistatics and electro-quasistatics, achieve their simplicity by neglecting the coupling between fields. Magneto-quasistatics truncates the Ampere-Maxwell Equation by neglecting the displacement current, $\partial_t(\varepsilon_0 \mathbf{E})$, and electro-quasistatics truncates Faraday's Law by neglecting the time derivative of the magnetic field, $\partial_t \mathbf{B}$. As a result, electro-quasistatics cannot describe inductive effects and magneto-quasistatics cannot describe capacitive effects. Each offers a one-sided view of physics and each directs a blind eye to the other side. Some situations involve both. For example, when placed in an oscillating magnetic field, a pair of wires – each twisted around the other – exhibits both capacitive and inductive effects. Attractively, ATM includes both while dismissing radiation.

If ATM is considered for use as a comprehensive approximation and not only as an example of a different way to generalize Ampere's Equation, one must ask if the ATM equations are tractable.

We already know that the Biot-Savart Law will provide the ATM magnetic field. And so we consider the electric field. Its evaluation requires both the scalar potential and the time derivative of the vector potential, $\mathbf{E} = -\nabla \varphi - \partial_t \mathbf{A}$. Unlike electro-quasistatics, ATM does not neglect the time derivative of the vector potential. Two plausible approaches to evaluation present themselves; work in either the Biot-Savart Gauge or the Coulomb Gauge. The former's vector potential depends only on the current density but its scalar potential depends on both the charge density and its second derivative with respect to time. (See Eq.IV.3) The latter's scalar potential depends only on the charge density but its vector potential depends on both the current density and the time derivative of the charge density. (See Eq.IV.4) Each approach brings us to a computational challenge. Indeed each brings us to the identical computational challenge because the both the gradient of scalar potential and the time derivative of the vector potential contribute to the electric field. The challenge is that the term depending on the time dependent charge density manifests itself in the convolution of two three dimensional integrals, each over all-space. If one works in the Biot-Savart Gauge, this term occurs in the scalar potential. If one works in the Coulomb Gauge, as we will, the challenge appears in the vector potential specifically in the second term on the right hand side of Eq.V.1

$$\mathbf{A}_C^{ATM}[t,\mathbf{r}] = \iiint_{all\ space} d^3r_s \frac{\mu_0 \mathbf{J}[t,\mathbf{r}_s]}{4\pi\|\mathbf{r}-\mathbf{r}_s\|} - \iiint_{all\ space} d^3z \frac{\mu_0 \nabla_z \iiint_{all\ space} \frac{d^3r_s}{4\pi} \frac{\partial_t \rho[t,\mathbf{r}_s]}{\|\mathbf{r}_s-\mathbf{z}\|}}{4\pi\|\mathbf{r}-\mathbf{z}\|} \qquad (V.1)$$

Its evaluation certainly appears more burdensome than anything required by the customary quasi-static approximations.

Happily, things are not as bad as they first look. A long calculation shows that the six dimensional integral can be reduced to a three dimensional integral with following result.[27]

$$\iiint_{all\ space} d^3z \frac{\mu_0 \nabla_z \iiint_{all\ space} \frac{d^3r_s}{4\pi} \frac{\partial_t \rho[t,\mathbf{r}_s]}{\|\mathbf{r}_s-\mathbf{z}\|}}{4\pi\|\mathbf{r}-\mathbf{z}\|} = \mu_0 \nabla_r \left\{ \iiint_{all\ space} \frac{d^3r_s}{4\pi} \mathbf{J}[t,\mathbf{r}_s] \cdot \frac{1}{2} \frac{\mathbf{r}-\mathbf{r}_s}{\|\mathbf{r}-\mathbf{r}_s\|} \right\} \qquad (V.2)$$

One sees that the six dimensional integral is a gradient and so its curl is zero. The six-dimensional integral does not affect the magnetic field, an observation that is wholly consistent with all we have said. Of course, its time derivative does contribute to the electric field; we need the time derivative of the whole vector potential. When the first and second terms of Eq.V.1 are added one finds

$$\mathbf{A}_C^{ATM}[t,\mathbf{r}] = \mu_0 \iiint_{all\ space} \frac{d^3r_s}{4\pi} \left\{ \frac{\mathbf{J}[t,\mathbf{r}_s] + \left(\mathbf{J}[t,\mathbf{r}_s] \cdot \hat{\Delta}_{rr_s}\right)\hat{\Delta}_{rr_s}}{2\|\mathbf{r}-\mathbf{r}_s\|} \right\} \qquad (V.3a)$$

where

$$\hat{\Delta}_{rr_s} \equiv \frac{\mathbf{r}-\mathbf{r}_s}{\|\mathbf{r}-\mathbf{r}_s\|} \qquad (V.3b)$$

One sees that despite its discouraging initial appearance, the Coulomb Gauge ATM vector potential is not qualitatively harder to evaluate than the Biot-Savart vector potential which is used in statics and magneto-quasistatics.

Another characteristic of Eq.V.3 deserves attention. The Coulomb Gauge ATM vector potential has the same form as the vector potential of Darwin's Approximation to the combination of relativistic particle mechanics and Maxwell's Equations, something we return to in the next section. [28]

**VI. History of some related matters**

No one can write about Maxwell's Equations without wondering if he has been anticipated or more likely, how many others have anticipated him. And so, the author began his search with the obviously relevant source, Maxwell's Treatise [29] to learn what the great man wrote about his displacement current and the alternatives he considered. Surprisingly, Maxwell wrote very little about them. In Article 607, one finds:

> "We have very little experimental evidence relating to the direct electromagnetic action of currents due to the variation of electric displacement in dielectrics, but the extreme difficulty of reconciling the laws of electromagnetism with the existence of electric currents which are not closed is one reason among many why we must admit the existence of transient currents due to the variation of displacement. Their importance will be seen when we come to the electro-magnetic theory of light."

And in his article 610, Maxwell wrote:

> "One of the chief peculiarities of this treatise is the doctrine which it asserts that the true electric current, that on which electromagnetic phenomena depend, is not the same thing as the current of conduction, but that the time variation of the electric displacement must be taken into account in estimating the total movement of electricity so that we must write…"

$$\mathbf{C} = \mathbf{J}_{conduction} + \partial_t \mathbf{D} \qquad \text{(Equation of True Currents)} \qquad \text{(H)}$$

While he offered little about his displacement current, Maxwell's characterization "…very little experimental evidence relating to…" did not hide anything; there was little to discuss. Indeed it has remained so. To the author's knowledge, M.R. van Cauwenberghe,[30] writing in 1929, was the first to report measuring the magnetic flux in a capacitor. And no further reports of such measurements were made until 1974 when T.R. Carver and J. Rahjel [31] described their apparatus and then until 1985 when D.F. Bartlett and T.R. Corle [32,33,34,35] measured the displacement current within a capacitor.

The paucity of measurement is matched by the apparent absence of proposed alternatives to Maxwell's displacement current. A review of some of the secondary literature [36,37,38,39,40,41,42] found no mention of charge conserving alternatives to Maxwell's displacement current, though it is reported that some of Maxwell's distinguished contemporaries simply did not accept it. Further, in gracious replies to this writer's queries, Prof. J.Z. Buchwald and Prof. D.M. Siegel have each said he is not aware of discussion, by either Maxwell or his contemporaries, of charge conserving alternatives to Maxwell's Equations.

The first appearance of work that in retrospect, and only in retrospect, bears on the topic of this article, was motivated by a completely different issue, the effect of the proton's motion on the electron. Stimulated by the work of Bohr and Sommerfeld on hydrogen, in 1920, C.G. Darwin [43] sought a way to take account of that motion. More generally, Darwin's concern was with particles and his focus was on slow relative motion, slow enough so that retardation could be neglected. By keeping the first variation of mass with velocity and by expanding the vector potential in powers of the retardation, he arrived at a Lagrangian that described instantaneous velocity dependent interaction among particles [44] and a velocity dependent mass, both correct to first order in v/c. Charge conservation was guaranteed by imputing a time independent charge to each point particle and assuming each was eternal. Subsequently, Darwin's work was adapted to the one particle Dirac Equation by G. Breit [45]. Much more recently, A.N. Kaufman and P.S. Rostler [46] proposed using Darwin's work to study plasmas without having to deal with radiation and others [47] have followed their suggestion. More recently, J. Larsson suggested that Darwin's approximation might be used to investigate eddy currents.[48]

The similarity and difference in approach between Darwin's pioneering work and that presented here can be easily stated. The similarity comes about because the alternative to the displacement current presented here involves spatially separated but simultaneous quantities and Darwin's approximation neglects retardation when calculating the currents' vector potential. The difference is that Darwin and followers refer to currents comprised of charged point particles while this formulation refers to any current, without regard to its constitution.

**VII. Summary & conclusions**

Contrary to statements in several otherwise exemplary textbooks, Maxwell's displacement current is not necessary to insure the local conservation of charge. Alternative generalizations of Ampere's Equation are possible and one has been exhibited here. The fields, via the potentials, were constructed from the charge and current densities. Physics students may take away the thought that, though a theory may be widely discussed, that theory may not be the only solution to the problem (e.g., incorporation of local charge conservation) that it was meant to overcome.

The crucial difference between Maxwell's displacement current and the alternative discussed here was shown to be this – the displacement current entails the propagation of electromagnetic influence at the speed of light while the alternative propagates influence instantaneously.

Because the finiteness of speed of light can be neglected when describing many phenomena, the possibility presents itself that, for them, the ATM may in fact be a good approximation to Maxwell. The ATM is conceptually superior to the usual magneto-quasistatic or electro-quasistatic approximations because each neglects an important aspect of the physics (either magnetic or electric energy) and so their users must close their eyes to situations that are neither purely inductive nor purely capacitive.

As with any candidate approximation, the question of tractability arises. We found that the solutions of the ATM Equations appear no more difficult to evaluate than the solutions of static problems. Further we noted that if one imagines the current to comprise classical moving charged point particles, one arrives at the Darwin Approximation, an approximation that was motivated by the wish to take account of semi-relativistic effects, not local charge conservation.

Approximation aside, ATM may stimulate to a better appreciation of Maxwell's theory. In that regard, we note that if Maxwell's displacement current is thought of as the sum of its longitudinal and transverse parts, then each of these two parts has its own distinct role. The longitudinal part conserves charge while the transverse part enables radiation to propagate at the speed of light in empty space. From that point of view, ATM keeps the longitudinal part of the displacement current while dismissing its transverse part.

**Acknowledgments**

The author is grateful to Professors Antonio Badía-Majós and Jesús Letosa for making possible his visit to Departamento de Física de la Materia Condensada of the Universidad de Zaragoza. Prof. Antonio Badía-Majós' probing questions stimulated improvements in an earlier version of this paper and are happily recalled, as is his company and that of each of the other friends made during that visit. The author is also grateful to Prof. Jed Z. Buchwald and Prof. Daniel M. Siegel, each of whom graciously replied to the author's queries about discussion by Maxwell's contemporaries and followers of charge conserving alternatives to Maxwell's displacement current. Finally, the author thanks two anonymous referees for drawing his attention to several references.

**References**

[1] James C. Maxwell, *A Treatise on Electricity and Magnetism, unabridged Third Edition (of 1891)*, vol.2, Dover Publications, Inc (New York, 1954) p. 253

[2] While preparing this paper, the author became aware that this point was made in H.S. Zapolsky, "Does charge conservation imply the displacement current?" Am. J. Phys., vol. 55, no. 12 (December, 1987) p. 1140 without however deducing any of its consequences, the topic of the rest of this paper.

[3] Edward M. Purcell, *Electricity and Magnetism, Berkeley Physics Course: volume* 2, 2$^{nd}$ edition, McGraw-Hill (Boston, Mass, 1985) p. 328

[4] Gerald L. Pollack and Daniel R. Stump, *Electromagnetism*, Addison-Wesley (San Francisco, 2002) page 404. Also see page 399 where the following appears "The modification consists of an additional term, called the *displacement current*, added to the right hand side, which we will derive presently. Remarkably, it is possible to construct the complete equation by a purely theoretical analysis, which is how it was found originally by Maxwell."

[5] Wolfgang Panofsky and Melba Phillips, *Classical Electricity and Magnetism*, Addison-Wesley (Reading, Mass, 1962) p. 136

[6] Walter Greiner, *Classical Electrodynamics*, Springer Verlag (New York, 1998) p. 251

[7] Anupam Garg, *Classical Electromagnetism in a Nutshell*, Princeton University Press (Princeton, 2012) p. 125

[8] David J. Griffiths, *Introduction to Electrodynamics*, 4$^{th}$ edition, Pearson (Boston, 2013) p. 335

[9] Functions of the form $\varphi[\mathbf{x}] = \left(e/4\pi\varepsilon_0 \|\mathbf{x}-\mathbf{q}\|\right) + e\sum_{lm} \|\mathbf{x}-\mathbf{q}\|^l c_{lm} Y_{lm}$ solve Poisson's Equation, $\nabla_x^2 \varphi[\mathbf{x}] = -e\delta^3[\mathbf{x}-\mathbf{q}]$, but do not satisfy the boundary condition at infinity, unless each of the constants, $c_{lm}$, is zero.

[10] Indeed, the author cheerfully admits that he long ago forgot this and that his own path to Eq.III.1 was roundabout - involving both Helmholtz' Theorem and several integrations by parts – not something to expound in 4,000 words.

[11] John D. Jackson, *Classical Electrodynamics*, third edition, Wiley (New York, 1999) p. 178.

[12] Wolfgang Panofsky and Melba Phillips, *Classical Electricity and Magnetism*, second edition Addison-Wesley (Reading, Mass, 1962) p. 128.

[13] T. Charitat and F. Graner, "About the magnetic field of a finite wire", Eur. J. Phys. vol. 24 (2003) pps. 267–270

[14] Oleg Jefimenko, *Electricity and Magnetism*, second edition, Electret Scientific Company (Star City, 1989) page 516

[15] David J. Griffiths and Mark A. Heald, "Time-dependent generalizations of the Biot-Savart and Coulomb laws", Am. J. Phys., vol. 59, no. 2 (February, 1991) pps. 111-117.

[16] Many other gauges are possible, for example the Multipolar Gauge and the Lorenz Gauge (formerly known as the Lorentz Gauge) though the latter is particularly cumbersome in the ATM.

[17] O.L. Brill and B. Goodman, "Causality in the Coulomb Gauge", Am. J. Phys., vol. 55, no. 9 (September, 1967) pps. 832-837.

[18] C.W. Gardiner and P.D. Drummond, ''Causality in the Coulomb gauge: A direct proof,'' Phys. Rev. A vol.38, (November,1988) pps. 4897–4898.

[19] B.J. Wundt and U.D. Jentschura, "Sources, Potentials and Fields in Lorenz and Coulomb Gauge: Cancellation of Instantaneous Interations for Moving Point Charges" arXiv: 1110.6210v3 [physics.class-ph] (9 September 2014) pps. 8.

[20] Andor Frenkel, and István Rácz, "On the use of projection operators in electrodynamics" European Journal of Physics vol.36 (2015) pps1-12.

[21] John D. Jackson, "Maxwell's displacement current revisited", Eur. J. Phys. vol.19 (1999) pps 495-499.

[22] Fritz Rohrlich, "Causality, the Coulomb Field, and Newton's law of gravitation", Am. J. Phys., vol. 70, no. 4 (April, 2002) pps. 411-414.

[23] Fritz Rohrlich, "The validity of the Helmholtz theorem", Am. J. Phys., vol. 72, no. 3 (March, 2004) pps. 412-413.

[24] José A. Heras, ''Comment on 'Causality, the Coulomb field, and Newton's law of gravitation,' '' by F. Rohrlich, Am. J. Phys. vol. 71, no.7, (July, 2003) pps.729 – 730.

[25] José A. Heras, "A formal interpretation of the displacement current and the instantaneous formulation of Maxwell's equations", Am. J. Phys., vol. 79, no. 4 (April, 2011) pps. 409-416.

[26] Kuo Ho Yang, "The physics of gauge transformations", Am. J. Phys., vol. 73, no. 8 (August, 2005) pps. 742-751.

[27] The calculation is presented and discussed in Alan M. Wolsky, "How a charge conserving alternative to Maxwell's displacement current entails a Darwin-like approximation to the solutions of Maxwell's equations", arXiv: 1411.7428 [physics.class-ph] (27 November 2014) pps. 8.

[28] John D. Jackson, *Classical Electrodynamics*, third edition, Wiley (New York, 1999) pps. 596-598

[29] James C. Maxwell, *A Treatise on Electricity and Magnetism, unabridged Third Edition (of 1891),* vol.2, Dover Publications, Inc (New York, 1954) pps 252 and 253

[30] M.R. van Cauwenberghe, *Vérification Expérimentale de l'équivalence Électromagnétique entre les Courants de Déplacement de Maxwell*, Journal de Physique (August, 1929) Number 8 pps. 303-312

[31] Thomas R. Carver and Jan Rajhel, "Direct "Literal" Demonstration of the Effect of a Displacement Current", Am. J. Phys. 42, 246 (March, 1974) pps. 246-249

[32] D. F. Bartlett and T. R. Corle, "Measuring Maxwell's Displacement Current Inside a Capacitor", Phys. Rev. Letters, vol. 55, no. 1, (1 July 1985) pps. 59-62

[33] John Maddox, "Measuring the Unmeasurable*"*, Nature vol.316, (11 July 1985) p. 101

[34] D. F. Bartlett, "Modest Disclaimer*"*, Nature vol.316, (29 August 1985) p. 760

[35] D. F. Bartlett and G. Gengul, "Measurement of Quasistatic Maxwell's Displacement Current*"*, Phys. Rev. A vol.39 (1989) pps. 938-945

[36] Edmund T. Whittaker, *A History of the Theories of Aether and Electricity: The Classical Theories/The modern theories*, 2 volumes, Harper and Brothers (New York, 1960)

[37] Alfred M. Bork, *Maxwell, "*Displacement Current, and Symmetry*"*, Am. J. Phys., vol. 31, (November, 1963) pps. 854-859

[38] Joan Bromberg, "Maxwell's Electrostatics", Am. J. Phys, vol. 36, no.2 (February, 1968) pps. 142-151

[39] Joan Bromberg 1967–8 "Maxwell's displacement current and his theory of light*"*, Arch. Hist. Exact Sci. 4 pps. 218–234

[40] Jed Z. Buchwald, *From Maxwell to Microphysics*, University of Chicago Press (Chicago, 1985)

[41] Daniel M. Siegel, *Innovation in Maxwell's electromagnetic theory, Molecular vortices, displacement current, and light*, Cambridge University Press (Cambridge, 1991)

[42] Edmund T. Whittaker, *A History of the Theories of Aether and Electricity: The Classical Theories/The modern theories,* Harper and Brothers (New York, 1960) see vol.1 pps. 266-270

[43] Charles G. Darwin, "The dynamical motions of charged particles", Phil. Mag. ser. 6. 39, pps.537-551 (1920)

[44] One might say that this part of the work was anticipated in Oliver Heaviside, "On the Electromagnetic Effects due to Motion of Electrification through a Dielectric*"*, The Philosophical Magazine 1889 pps. 324-339

[45] Gregory Breit, "The Effect of Retardation on the Interaction of Two Electrons", Phys. Rev. vol.34, no.4 (August 15, 1929) pps. 553-573

[46] Allan N. Kaufman and Peter S. Rostler, "The Darwin Model as a Tool for Electromagnetic Plasma Simulation", Phys. Fluids **14**, (1971) pps 446-448

[47] For example, see Todd B. Krause, Amit Apte, and Philip J. Morrison, "A unified approach to the Darwin approximation", Physics of Plasmas, vol.14 (2007) p. 102112 and references cited therein.

[48] Jonas Larsson, "Electromagnetics from a quasistatic perspective", Am. J. Phys, vol.75, no. 3, (March, 2007) pps. 230-239.